# Global Albedos of Pluto and Charon from LORRI *New Horizons* Observations

B. J. Buratti[1], J. D. Hofgartner[1], M. D. Hicks[1], H. A. Weaver[2], S. A. Stern[3], T. Momary[1], J. A. Mosher[1], R. A. Beyer[4], A. J. Verbiscer[5], A. M. Zangari[3], L. A. Young[3], C. M. Lisse[2], K. Singer[3], A. Cheng[2], W. Grundy[6], K. Ennico[4], C. B. Olkin[3]

[1]Jet Propulsion Laboratory, California Institute of Technology, Pasadena, CA 91109, bonnie.buratti@jpl.nasa.gov
[2]Johns Hopkins University Applied Physics Laboratory, Laurel, MD 20723
[3]Southwest Research Institute, Boulder, CO 80302
[4]National Aeronautics and Space Administration (NASA) Ames Research Center, Moffett Field, CA 94035
[5]University of Virginia, Charlottesville, VA 22904
[6]Lowell Observatory, Flagstaff, AZ

Keywords: Pluto; Charon; surfaces; New Horizons, LORRI; Kuiper Belt

32 Pages

3 Tables

8 Figures

**Abstract**

The exploration of the Pluto-Charon system by the *New Horizons* spacecraft represents the first opportunity to understand the distribution of albedo and other photometric properties of the surfaces of objects in the Solar System's "Third Zone" of distant ice-rich bodies. Images of the entire illuminated surface of Pluto and Charon obtained by the Long Range Reconnaissance Imager (LORRI) camera provide a global map of Pluto that reveals surface albedo variegations larger than any other Solar System world except for Saturn's moon Iapetus. Normal reflectances on Pluto range from 0.08-1.0, and the low-albedo areas of Pluto are darker than any region of Charon. Charon exhibits a much blander surface with normal reflectances ranging from 0.20-0.73. Pluto's albedo features are well-correlated with geologic features, although some exogenous low-albedo dust may be responsible for features seen to the west of the area informally named Tombaugh Regio. The albedo patterns of both Pluto and Charon are latitudinally organized, with the exception of Tombaugh Regio, with darker regions concentrated at the Pluto's equator and Charon's northern pole The phase curve of Pluto is similar to that of Triton, the large moon of Neptune believed to be a captured Kuiper Belt Object (KBO), while Charon's is similar to that of the Moon. Preliminary Bond albedos are 0.25±0.03 for Charon and 0.72±0.07 for Pluto. Maps of an approximation to the Bond albedo for both Pluto and Charon are presented for the first time. Our work shows a connection between very high albedo (near unity) and planetary activity, a result that suggests the KBO Eris may be currently active.

**Introduction**

Quantitative measurements of the albedo of planetary surfaces yield clues to geological processes, including resurfacing, exogenous alterations by meteoritic impact or accretion of dust, magnetospheric interactions, and bombardment by ionizing photons. Observations by the Long Range Reconnaissance Imager (LORRI) camera on the *New Horizons* spacecraft offer the first global, highly resolved measurements of dwarf planet Pluto, its companion Charon, and four minor moons, the first system in the Solar System's "Third Zone" to be visited by a spacecraft (see Cheng et al., 2008 for a description of the camera). This region is populated by small, ice-rich bodies that are distinct from the gas giants and the rocky terrestrial planets. Prior to the spacecraft's closest approach LORRI obtained views of the global albedo variations on Pluto– the focus of this paper – while during closest approach the spacecraft imaged the surface at sub-km resolution to provide a view of albedo patterns within the context of geologic features and exogenous alteration processes.

Ground-based observations revealed large albedo variations on Pluto. A lightcurve of about 0.3 magnitudes in the blue and visible region of the spectrum and albedo maps based on Pluto-Charon mutual events both suggested high-albedo regions juxtaposed to much lower albedo areas (Stern et al., 1997; Buie et al. 2010a,b; Buratti et al., 2003; 2015). The lightcurve pattern was not sinusoidal such as those of the Saturnian moons, and to a lesser extent the three outer Galilean moons, which exhibit albedo patterns largely due to exogenous processes (Johnson et al., 1983; Buratti et al., 1990; Verbiscer et al. 2007; Schenk et al. 2011). Iapetus exhibits the largest albedo variations on any airless body, more than a factor of 10. These variations are due almost entirely to accretion of low-albedo dust from Saturn's Phoebe ring, augmented by thermal migration (Buratti and Mosher 1995; Verbiscer et al., 2009; Spencer and Denk, 2010).

Dione has a lightcurve of nearly 0.4 mag, and albedo variegations of at least a factor of two (Buratti, 1984; Buratti and Veverka, 1984), but it is almost all due to exogenous processes such as accretion of E-ring particles, and magnetospheric and meteoritic bombardment (Buratti et al., 1990; Schenk et al., 2011) possibly augmented by thermal migration (Blackburn et al. 2012). For Pluto a model with two spots separated by 134° in longitude and with albedos twice that of the surrounding terrain, which could likely exist as a low-albedo longitudinal band, explained the photoelectric lightcurves measured between 1954 and 1988 (Marcialis, 1988). Although not unique, this model showed an early awareness of stark albedo differences on Pluto's surface, including the possibility of "polar caps with albedos near unity". The production of a map from the mutual event season pinpointed a very bright localized feature "that may be due to condensation around a geyser or in a crater" (Young et al., 1999). Finally, Stern et al. (1988) pointed out that the replenishment of seasonal volatiles on a periodic basis would lead to high albedos.

Since the turn of the millennium, Pluto also showed changes in its lightcurve beyond those expected for a static frost model in which the only temporal variations in albedo are those due to the easily calculated excursions in the radiance angles (Buratti et al., 2015). *Hubble Space Telescope* (*HST*) maps obtained in 2002 (Buie et al., 2010b) also showed slight changes in albedo that were consistent with those of the rotational lightcurves, both in terms of the area undergoing changes and the amount of the change. Pluto seemed to join Triton as an icy body in the outer Solar System that was undergoing seasonal volatile transport on its surface (Bauer et al., 2010; Buratti et al., 2011), with the possibility of active geologic processes being responsible for the changes as well.

In remote observations Charon exhibited much smaller albedo variations than Pluto (Buie et al., 2010a, b), suggesting a far different and less complex history. No changes through time were observed in Charon's lightcurve or on its surface as imaged with the *HST*. Thus, all ground-based photometric measurements obtained prior to the *New Horizons* encounter with the Pluto system suggested these two worlds were very different, with Pluto being the more dynamic of the two.

Most variations in the specific intensity of a planetary surface are not intrinsic, but rather due to changes in the incident, emission, and solar phase angles. The variations in incident and emission angles, often called the photometric function, need to be modeled and fully accounted for to produce a map of the intrinsic reflectivity of a surface. Additional changes in the intensity are also due to factors that are a function of the physical nature of the surface, including macroscopic roughness, which alters the local incident and emission angles of the surface and removes radiation through shadowing; non-isotropy in the single particle phase function; and mutual shadowing among the small particles comprising the optically active portion of the regolith. The latter effect, which is responsible for the opposition surge observed on Pluto (Buratti et al., 2015), along with other effects such as coherent backscatter, cannot be studied by *New Horizons* because it never reached the small (<6°) solar phase angles necessary to characterize the surge. Verbiscer et al. (2016) provide the opposition surge observations for Pluto and Charon from HST during the New Horizons epoch at phase angles ranging from 0.06-1.72°.

The goal of this paper is to derive global normal reflectances for Pluto and Charon with all geometric effects removed, and to produce a preliminary map of the Bond albedo for both objects. The latter is an integral part of thermal models for the surface of these objects, and for understanding energy balance on them. We focus on global albedo patterns, with some first analyses of disk-resolved images of Pluto, with emphasis on quantitative albedo differences on the surface and their connection with the underlying geology.

The approach images of Pluto and Charon were obtained at relatively small solar phase angles: 11°-17° (with the ones in this paper covering 15°-17°), obviating the necessity of fitting complex photometric models and extrapolating them to normal reflectances. Radiative transfer models, which connect the intensity to the physical properties of the surface (Hapke 1981; 1984; 1986; 1990; Goguen, 1981;Shkuratov et al., 2005; Irvine, 1966; Buratti, 1985), require a full excursion in viewing geometry and simultaneous analysis of disk-resolved and disk-integrated observations to derive unique information (Helfenstein et al, 1988). We instead take a more empirical approach which seeks to utilize ground based and *New Horizons* measurements of Pluto's and Charon's solar phase curve to extrapolate the approach images to normal reflectances.

All names of features used in this study are informal ones as presented in Stern et al. (2015) that have been adopted by the *New Horizons* flight team to facilitate uniform discussion and analysis. They have not been approved by the International Astronomical Union (IAU).

**Observations and Data Analysis**

Pluto and Charon are tidally evolved such that they rotate about a common center of gravity every 6.387 days. In the week leading up to the *New Horizons* flyby of Pluto and its moon, images at all longitudes of Pluto were obtained by the remote sensing instruments on the spacecraft. (Closest approach images provided, of course, disk-resolved measurements of only one hemisphere of Pluto and Charon). LORRI images obtained in the week leading up to the *New Horizons* closest approach of Pluto thus provide a global map of all the illuminated regions of Pluto's and Charon's disk.

The LORRI images used for constructing the maps of normal reflectance of Pluto and Charon are listed in Table. 1, along with their integration times and their associated geometric information including solar phase angle, range, subspaccraft and subsolar geographical latitude and longitude, and spatial resolution These images represent the best spatial resolution obtained for each geographical location within the week prior to closest approach. For most of the data, Pluto and Charon appear on the same image (It wasn't until three days before closest approach that the binary pair exceeded the LORRI Field-of-View.) Pipeline calibration procedures were employed to flatfield each image, remove blemishes, and transform data numbers (DNs) into radiometric units using the flight calibration current as of late February 2016.

*Global maps of normal reflectance*

Since geologic analysis of images requires the knowledge of intrinsic values of the albedo, changes due solely to viewing geometry must be modeled and removed from the data. The images used in this study were obtained at small solar phase angles (although still larger than any observed from Earth); thus the corrections for solar phase angle effects are not large.

Photometric changes on a surface are due to two primary factors: changes in the viewing geometry as the incident, emission, and solar phase angle change, and the physical character of the surface. This latter factor includes the anisotropy of scatterings in the surface, which is expressed by the single particle phase function; the compaction state of the surface, which leads to the well-known opposition surge attributed to the rapid disappearance of mutual shadows among regolith particles as the surface becomes fully illuminated to an observer, and to coherent backscatter (Irvine, 1966; Hapke, 1990); and to macroscopic roughness, which both alters the local incident and emission angles and removes radiation due to shadowing (Hapke, 1984; Buratti and Veverka, 1985). Radiative transfer models have been developed that fully describe the specific intensity returned from a planetary surface (Horak 1950; Chandrasekhar, 1960; Goguen, 1981; Hapke, 1981; 1984; 1986; 1990; Buratti, 1985; Shkuratov, 2005). These models suffer from a number of shortcomings, foremost among them are that they do not represent physical reality well, and that even with the complete data sets returned by spacecraft, unique fits to physical parameters cannot be made (Buratti, 1985; Helfenstein et al., 1988; Shepard and Helfenstein, 2007). The latter problem is particularly acute with a planetary flyby, and in the case of *New Horizons*, much of the data at large solar phase angles need to be corrected for atmospheric contributions, as was done for Triton (Hillier et al. 1990). Some recent work by Shephard and Helfenstein (2011) and Helfenstein and Shepard (2011) has been more positive, showing that unique fits of physical photometric parameters can be made to surfaces of low to moderate albedos.

Fortunately, simple empirical photometric models have been developed that are more appropriate for the data set in hand: observations at small solar phase angles (~10°-15°) leading up to the flyby. Two widely used models are those of Minnaert (1961), which is essentially a first-order Fourier fit that describes the distribution of intensity on a planetary surface, and a lunar-Lambert model that is the superposition of a lunar, or Lommel-Seeliger law, describing singly scattered radiation, and a Lambert law describing multiple scattered photons (Squyres and Veverka, 1981; Buratti, 1984):

$$\frac{I}{F} = \frac{f(\alpha) A \mu_0}{\mu_0 + \mu} + (1 - A)\mu_0 \qquad (1)$$

where I is the intensity of scattered light at a point on a planetary disk, $\pi F$ is the incident solar flux at that point, $\alpha$ is the solar phase angle, $\mu_o$ is the cosine of the incident angle, $\mu$ is the cosine of the emission angle, and A is the fraction of radiation that is singly scattered. The term $f(\alpha)$ is the surface phase function: it expresses changes on the surface due to the physical properties of roughness, compaction state, and scattering anisotropy defined above. This equation is only semi-empirical, as it contains the leading terms of more complete equations of radiative transfer (Goguen, 1981).

This paper presents global maps for two fundamental photometric properties for Pluto and Charon: the normal reflectance and the Bond albedo. The first quantity expresses the I/F for incident angle, emission angle, and $\alpha=0$. For the photometric function defined by eq. (1) the normal reflectance is:

$$r_n = \left(\frac{I}{F}\right)_{measured} \frac{\frac{f(0°)A}{2} + (1-A)}{\frac{f(\alpha)A\mu_0}{\mu_0 + \mu} + (1-A)\mu_0} \qquad (2)$$

The surface phase function $f(\alpha)$ can be computed by using equations derived in Buratti and Veverka (1983; 1984) based on integral solar phase curves, or it can be measured from an image or from a point on the surface. This function at 0° can be derived by fitting a function to $f(\alpha)$ measured at larger solar phase angles, or it can be derived from the following equation (Buratti and Veverka, 1983):

$$p=2/3(1-A) + A(f(0°))/2 \qquad (3)$$

where p is the geometric albedo, defined as the brightness of an object at a solar phase angle of 0° relative to a perfectly diffuse (Lambert) disk of the same size also at 0°. Note that this equation "partitions" the geometric albedo between the Lambert portion of the photometric function, for which the geometric albedo is 2/3 the normal reflectance, and the Lommel-Seeliger portion, for which f(0°) equals twice the geometric albedo. (The normal reflectance equals the geometric albedo for a Lommel-Seeliger surface). The geometric albedo of Pluto close to the *New Horizons* encounter is given in Buratti et al. (2015) as 0.62±0.03 at 0.62µm, which is near the pivot wavelength of the LORRI camera of 0.607µm (the pivot wavelength is close to the effective wavelength for a spectrally flat source; for an exact definition see Cheng et al. 2008). A is derived through a best fit procedure.

To obtain integral solar phase curves of Pluto and Charon, we applied the LORRI flight calibration current as of late February 2016 to images at the range of solar phase angles for which full-disk images were available. Integral photometric procedures using IRAF were applied to these images, which were corrected for rotational light variations and distance from the *New Horizons* spacecraft using best-fit procedures described in Buratti et al. (2015). There are observations at very large solar phases which were not included because of scattered light problems (Charon) and atmospheric contamination (Pluto). A series of images averaged over a rotational light curve at 11° and 14.5° was used for Pluto, while a similar set of images was used for Charon at 14.5°. For the zero-point of the phase curve (which was not observed from *New Horizons*), we used the ground-based values for both Pluto and Charon of Buie et al. (2010a), which are in good agreement with those of Buratti et al. (2015) for Pluto. Figure 1 shows the phase curves of Pluto and Charon in the V-filter compared with several icy moons, the Moon, and Mercury.

***Pluto.*** For Pluto the challenge is in assigning a photometric function to a surface that exhibits a wide range in terrains and albedos. Preliminary studies of Pluto with *New Horizons* measurements indicated the albedo variations on Pluto were large, as suggested by both ground-based light curves and *HST* maps (Young et al., 1999; Stern et al, 1997; Buie et al., 2010a,b; Buratti et al, 2015; Grundy et al. 2016a). These wide variations in albedo indicate that the scattering properties of Pluto are expected to change as a function of albedo and position on the body. This complicated situation cannot be fully disentangled, as in many cases changes due to excursions in incident and emission angle mimic changes in albedo. This problem is especially acute near the polar cap(s) of Pluto where the increasing thickness of frost is associated with a gradual increase in albedo. For this first study we shall adopt a single photometric function for Pluto that represents the best global fit to the surface of Pluto. A best-fit value of A=0.70 was found by seeking a photometric function that minimized offsets between overlapping images. The surface phase function at 0°, $f(0°)$, was calculated from eq. 3 and the ground-based geometric albedo (Buratti et al., 2015, Buie et al. 2010a). The normal reflectances were derived from eq. 2. The surface phase function for each solar phase angle, $f(\alpha)$, was computed using the observed disk-integrated solar phase function of Pluto (Figure 1) and eq. 2 on p. 403 of Buratti (1984). The use of a single photometric function for Pluto results in an underestimate of the highest albedo regions of Pluto, as the specific intensity is undercorrected for the effects of limb-darkening. Likewise, the lowest albedos are overestimated. Since the extreme albedo regions on our maps do not occur in regions with high incident angles, this effect is not great.

Figure 2 shows the map of normal reflectance for Pluto obtained with the images listed in Table 1. Normal reflectances range from a low of 0.08 in the lowest-albedo regions of Cthulhu Regio (between latitudes of -30°S to 0° and longitudes of 40°E to 170°E) to 1.0 in the highest albedo regions of the feature informally named Sputnik Planitia (between latitudes of 0° to 45°N and longitudes of 160°-200°). No other body except the Saturnian satellite Iapetus exhibits such large albedo variegations, and those are due to an exogenous process, augmented by thermal segregation (Buratti and Mosher, 1995; Spencer and Denk, 2010). Pluto also has a distinct polar cap in the north and there are hints of one in the south as well, but they are not nearly as bright as Tombaugh Regio, the large high-albedo region near the middle of the map.

***Charon.*** . The albedo of Charon is relatively uniform; the problem of the photometric function changing with position on its surface is thus avoided. Previous work on the photometric functions of icy moons suggested that Charon would exhibit scattering properties similar to that of the Moon. Squyres and Veverka (1981) found that the surface of Ganymede with a visible geometric albedo of 0.43, slightly higher than that of Charon, could be described by a lunar-like photometric function (A=1). But based on *Dawn* Framing Camera observations, Asteroid 4 Vesta with a visible geometric albedo of 0.43 had a small degree of non-lunar like scattering (Li et al., 2013). Buratti (1984) found that the icy moons of Saturn, with geometric albedos ranging from about 0.40 to 1.0, closely followed a lunar-like photometric function for geometric albedos of about 0.55-0.60. Charon was well within the range of icy surfaces following lunar-like scattering. Thus, it was surprising that the best-fit photometric function is similar to that of Pluto, with A=0.70. Figure 3 shows three maps of the geometric albedo of Charon with various values of A: 1.0; 0.7; and 0.5. Clearly, the value of 1.0 does not adequately describe the scattering properties of Charon.

Figure 4 shows the map of normal reflectance for Charon, processed in the same manner as that of Pluto. Charon does not exhibit the wide range in albedo that Pluto does: most of Charon's surface is characterized by normal reflectances in the 0.4-0.6 range, with a few bright crater ejecta areas. There is a substantially lower albedo polar region with normal reflectances of ~0.20. This region has been proposed as a cold trap for methane escaping from Pluto's atmosphere. This methane is photolyzed into more complex, lower albedo molecules during arctic winter and left behind as a lag deposit as more volatile pure methane sublimates with arrival of sunlight (Grundy et al., 2016b).

**The Bond Albedo**

The Bond albedo is a measure of the energy balance of a body: it is the ratio of the power at a specific wavelength on a body that is scattered back out into space. When integrated over the entire electromagnetic spectrum, it is the bolometric Bond albedo. More formally, the Bond albedo is equal to p, the geometric albedo, times q, the phase integral. The latter quantity, which expresses the directional scattering properties of a planetary body, can then be calculated with the following expression:

$$q(\lambda) = 2\int_0^\pi \Phi(\alpha,\lambda) \sin\alpha \, d\alpha \qquad (4)$$

where $\Phi(\lambda)$ is the disk-integrated normalized phase curve. The Bond albedo is a fundamental parameter for understanding energy balance and volatile transport on any planetary surface.

The Bond albedo is a disk-integrated quantity, viz., the geometric albedo times the phase integral, which are both disk-integrated parameters. Nevertheless, to do effective thermal modeling of specific regions on planetary surfaces, we map the quantity p*q to illustrate how the angle- integrated bidirectional reflectance varies over the surface. Such a map can be created by solving for the geometric albedo for a point source on the surface times the phase integral for that point. For low-albedo surfaces, the geometric albedo and normal reflectance – a disk resolved quantity - are equal, and for surfaces of arbitrary albedo, an equivalent geometric albedo is easy to derive (e.g., Buratti and Veverka, 1983). A phase integral for each point, based on the full range of geometric conditions applicable to that point, should be employed rather than a true phase integral. Squyres and Veverka (1982) show, for example, that the preponderance of lower incident angles near the poles and terminators will result in lower temperatures there. However, we do not have adequate observations from New Horizons for the appropriate excursion in viewing geometries for specific points, and given the changing albedo and corresponding photometric function for specific regions of Pluto's surface, computing such a "phase integral" is intractable. As an approximation, we use the disk-integrated phase integral. This approach has been useful in the past for computing a similar map for Iapetus (Blackburn et al. 2011).

A preliminary map of the Bond albedo at LORRI wavelengths can be constructed with a rudimentary phase curve and our normal albedo maps. LORRI Images of Pluto and Charon for which the full disk is included in the image exist for a small range of solar phase angles. As stated above, the images at large solar phase angles are contaminated by scattered light or atmospheric contributions in the case of Pluto. In future studies, synthetic integral values of Pluto's and Charon's solar phase curves will be constructed from disk-resolved observations. For these preliminary Bond albedo maps, we make use of the fact that phase integrals of objects that scatter like Pluto and Charon have been derived, and we use these values for this study. For Pluto we adopt the phase integral of Triton of 1.16 derived from *Voyager* images obtained in the green filter, which at 0.55μm is the closest in wavelength to LORRI (Hillier et al., 1990). For Charon, we use the lunar phase integral at 0.63μm of 0.60 (Lane and Irvine, 1972). Figure 1 shows that the integral phase function of Charon is similar to the Moon for the phase angles available. Thus for this preliminary study, the assumption of a lunar-like phase curve for Charon is reasonable.

For a lunar-like scattering function, the normal reflectance, which is a disk-resolved parameter, and the geometric albedo, which is a disk-integrated value, are equal. For Pluto and Charon, the geometric albedo can be derived from eq. 3, but deriving f(0°) involves extracting f($\alpha$) from a large number of solar phase angles and types of terrain and fitting a curve of f($\alpha$) to 0°, which was never observed with *New Horizons*. An easier and ultimately more accurate method of obtaining a geometric albedo is to normalize the maps in Figures 2 and 4 to geometric albedos determined from ground based observations. For Pluto, the value is 0.62±0.02 near the time of the New Horizons encounter for the R-filter at Table Mountain Observatory, which is centered at 0.62μm (Buratti et al., 2015), near the LORRI pivot wavelength of 0.607, while for Charon, it can be computed from the *New Horizons* radius of 606 km (Stern et al., 2015) combined with the ground-based opposition magnitude of 17.10 (Buie et al., 2010a), transformed to the R-filter using the spectrum of Charon (Sawyer et al., 1987; Fink and Disanti, 1988). This method yields a geometric albedo at LORRI wavelengths of 0.41±0.01. These normalized maps were multiplied by the phase integrals for Triton (in the case of Pluto) and the Moon (for the case of Charon). The preliminary Bond albedo of Pluto is 0.72±0.07 and that of Charon is 0.25±0.03

Figures 5 and 6 show the preliminary maps of the Bond albedo for both Pluto and Charon. Again, Pluto has wide albedo variegations, leading to large temperature changes and thermal segregation of volatiles on the surface. The low-albedo equatorial regions have Bond albedos in the 0.1-0.2 range; the actual values are even lower than those of this preliminary map, as when phase integrals defined by various regions of Pluto are known, the low-albedo regions will have smaller phase integrals than the disk-averaged value we used. Charon on the other hand, exhibits a small range of Bond albedos in the 0.2-0.3 range for most of its surface. Only the Bond albedo of its polar region is substantially lower.

**High resolution albedo maps**

In addition to constructing global maps of normal reflectance and Bond albedo, we have selected a sample of images to understand albedo variegations on smaller scales and to put the results in the context of Pluto's and Charon's geologic features that were revealed during the close encounter. Images at the interfaces between low- and high albedo regions are especially valuable. Among the questions that can be answered by these maps of intrinsic reflectivity from which all the effects of viewing geometry have been removed include:

1. How are the albedo changes correlated with geologic features, compositional units, and exogenous deposits?

2. What is the nature of Pluto's low-albedo material? Are there different types of low-albedo material? How is this material related to other dark material in the outer Solar System, such as that on Iapetus, Hyperion, Phoebe, and the Uranian moons?

3. Is the material on Pluto bi-modally distributed in albedo or is there a continuum?

4. What is the nature of the frost deposits on Pluto and how are they related to changes observed in ground-based measurements (Buie et al., 2010a,b; Buratti et al., 2015)?

5. Is albedo correlated to crater counts? Can albedo be used as a proxy for age?

6. Do crater ejecta materials exhibit any variation in albedo?

Table 2 lists images of Pluto and Charon obtained in the close-approach period that were transformed to normal reflectances with the same techniques as those used for the global maps. Figure 7 is a map of Pluto with the locations of these images and Figure 8 shows these five images of Pluto and one of Charon converted to normal reflectances. Along with the images are histograms of the frequency each value of normal reflectance occurs in each image. For Pluto, the immense variations in albedo are correlated with geologic features. For images that span both Tombaugh Regio and Cthulhu Regio, the two terrains are clearly bifurcated, with Cthulhu exhibiting reflectances of 0.1-0.2 and Tombaugh Regio exhibiting normal reflectances of 0.8-1.0. The only other body in the Solar System with such albedo extremes is Iapetus (see Table 3), but its variations are due to an exogenous deposit rather than geologic processes (although subsequent thermal segregation probably played a role in accentuating the albedo variations; see Spencer and Denk, 2010).

There are at least two distinct types of low-albedo terrain on Pluto. The first is represented by the equatorial band; the largest segment has been informally named Cthulhu Regio. It appears with the lowest normal reflectances of 0.08-0.2 in four out of five of the LORRI images of Pluto. There is another distinct type of low-albedo material that appears most clearly in the upper middle image of Figure 8 (LOR_0299177051). The upper right quadrant of this image is dominated by material that is morphologically similar to the "black rain" seen on Callisto, which Bottke et al. (2013) attribute to accretion of dust similar to that seen on Iapetus. Bottke et al. claim that as on Iapetus, Callisto's deposit is an accumulation of dust from a large outer ring. The normal reflectances of Pluto's possible "black rain" are distinctly different from Cthulhu Regio, in the range of 0.2-0.3. This bifurcation is most easily seen in the third image (LOR_0299177087), where the lower albedo characteristic of Cthulhu in the lower left region of the image, and the higher albedo associated with the "black rain" in the upper right portion of the image, each have their own bumps in the histogram. There is additional low-albedo material in the lower right part of the image which may be dune material or other aeolian deposits.

The idea of "black rain" on Pluto has several problems. First, why does it not exist on Charon as well? Even though Pluto acts as the main gravity well in the system, some exogenous material should be accreted onto Pluto's main moon. The low albedo deposits may instead be particles formed in the haze-layers of Pluto's atmosphere, or they may simply be low-albedo terrain similar to Cthulhu overlain with brighter material which increases its albedo.

The high-albedo terrain also exhibits a bifurcation. In the first image, there are two distinct regions, one centered on a normal reflectance of 0.90, and the other with a normal reflectance of 0.95. These distinct values suggest different episodes of resurfacing on Sputnik Planitia, which in fact appear on the bright terrain (lower right of first image, LOR_0299168135). Both albedos are extraordinarily high and are consistent with fresh ice or snow and evidence of recent activity. Bright regions close to Tombaugh Regio possess albedos close to that feature (see fourth figure, LOR_0299177195) implying they are also very fresh deposits of frost or snow, or that they were originally part of the same geologic unit, part of which has sublimated away. There is also an intermediate terrain with a broader distribution centered at normal reflectances of ~0.5. These areas could be a mechanical mixture of the Cthulhu Regio and Sputnik Planitia units, but they are more likely a separate unit, possibly even the edge of the polar cap.

Charon possesses three distinct albedo regimes. The first is a low albedo region corresponding to the dark feature at the pole with normal reflectances in the 0.20-0.35 range; this region is also redder than the terrain in the equatorial regions (Stern et al., 2015, Grundy et al. 2016b). Surrounding this lowest albedo region is an area of intermediate normal reflectances in the 0.35-0.45 range. These albedo features may be due to an exogenously placed deposit, or they are the result of an impact or a tectonic event that uncovered a deep layer of different composition. If the low-albedo feature is an impact basin, with the distinctly darkest material is the floor of the basin representing an excavated lower layer of different composition, the surrounding region could be ejecta deposit mixed with preexisting surface material. The alternate theory mentioned above (Grundy et al., 2016b) that involves an origin from Pluto's atmosphere and subsequent chemical alteration explains the albedo as well as the color of this feature, which is informally named Mordor. Although the geometric albedo of Charon is lower than that of Pluto, it does not possess the albedo range and variety that Pluto does, which is characteristic of its more quiescent history.

**Discussion and conclusions**

Pluto possesses an extraordinary range in albedo. Table 3 gives a quantitative indication of how Pluto compares with other objects in terms of its albedo variations: in a word, it is extreme, surpassed only by Iapetus. (Values for the albedo cited from our own work correspond to normal reflectances, but the other sources quote albedos at small solar phase angles.) Triton, which is believed to be a captured KBO (Agnor and Hamilton 2006), and which was thought to be an analogue to Pluto prior to the *New Horizons* flyby, exhibits albedo variations of about 50%, and it lacks very low-albedo terrain. Furthermore, the large ranges in albedo on other icy bodies are due to exogenous alterations such as accretion of low-albedo dust (Iapetus; Buratti and Mosher, 1995) or micrometeoritic and magnetospheric bombardment and accretion of bright particles from Saturn's E-ring (Dione; Buratti et al. 1990). Pluto's range in albedo can be explained by an extraordinary variety of ongoing geologic processes (Stern et al., 2015), with potential minor variations due to patterns of local insolation and exogenous dust deposition, including haze-particle settling. Albedo differences on Iapetus have been maintained and accentuated by thermal segregation (Spencer and Denk, 2010), a mechanism that may be at work on Pluto as well. With normal reflectances of 0.08 to nearly 1.0, which correspond to geometric albedos of 0.08 and a number somewhat less than 1.0 (as stated in the previous section, because of limb darkening at a solar phase angle of 0°, the geometric albedo of a surface with a Lambert component is less than its normal reflectance; for a pure Lambert scatterer it is 0.67 of the normal reflectance). With all other factors being equal (primarily emissivity and the phase integral), these albedo differences correspond to temperature differences of up to 20 K. But in reality, the phase integral of low-albedo regions is typically much lower: the temperature variations and corresponding cold trapping on Tombaugh Regio and clearing of Cthulhu Regio must be very efficient. The amount of energy absorbed by a planetary surface is 1-the Bond albedo (or more correctly the hemispheric albedo). Given the high Bond albedos in Tombaugh Regio, this small difference means it is very cold there. But given that the temperature is low even in low-albedo regions, due to the small amount of incident sunlight, the comparable fraction of energy absorbed in low and high albedo regions varies widely. Additional disk-resolved photometric analysis will quantify these differences and will provide a foundation for understanding the transport of volatiles on Pluto's surface.

We note that our maps of the Bond albedo from this preliminary study are only a first step in the creation of data products for this important physical parameter. The large albedo variegations on Pluto mean the presence of large temperature differences which may drive some of the active processes seen on the surface, and there must be some means of quantifying a regional energy balance when only limited observations are available. Even more essential than taking account of viewing geometry is modeling the different photometric and surface phase functions that occur because of the huge albedo variations on Pluto's surface. One downside of the LORRI data is that there are no disk-integrated data beyond $\alpha \sim 16°$. Equivalent disk-integrated brightness can be determined by computing the surface phase function $f(\alpha)$ and constructing a sphere of equivalent scattering properties (Buratti and Veverka, 1983; Buratti, 1984). Future work will focus on creating these brightnesses for both low-albedo and high-albedo regions of Pluto, and applying these more realistic phase integrals to specific regions.

Comparison of Pluto's albedo markings with those of Triton reveals how different these two icy worlds are. Like Triton, Pluto has at least one high-albedo polar cap, but Triton lacks any analogue to the very high albedo Tombaugh Regio. A better analogy might be the south polar active region (the "tiger stripes") of Enceladus, with comparable albedos (Table 3). Both the tiger stripes and Tombaugh Regio are regions of ongoing activity: active cryovolcanism on Enceladus and corresponding deposits of fresh plume particles, and glaciation with condensation due to cold trapping of volatiles such as methane for Tombaugh Regio. Composition maps of Pluto show an enhanced abundance of methane, nitrogen, and CO at Tombaugh Regio and methane in Pluto's pole, and a depletion of volatiles in the low-albedo Cthulhu Regio (Grundy et al. 2016a)

The seasonal transport of frost on Triton was detected from the ground (Bauer et al., 2010; Buratti et al. 2011), and what was thought to be seasonal volatile transport was observed on Pluto in the 60-year period between 1954 and 2013 (Buratti et al. 2015). *Hubble Space Telescope* maps also show albedo changes with time (Buie et al. 2010b). The regions of albedo change in Pluto's lightcurve – subobserver longitudes of ~140°-300° - correspond to the location of Tombaugh Regio. The ground-based observations suggest the removal of volatiles from the edges of this region. Moreover, the "reddening' of Pluto observed in the lightcurves after 2000 (Buie et al. 2010a,b; Buratti et al., 2015), particularly in the region of Cthulhu Regio, which is the reddest region of Pluto (Stern et al., 2015), also imply the removal of volatiles.

The albedo patterns on Pluto and Charon are both organized latitudinally (with the exception of Tombaugh Regio). Binzel et al (2016) explain these patterns in terms of insolation patterns forming polar, temperate, and tropical zones with corresponding degrees of volatile persistence. Future work that describes photometric functions for disk-resolved regions of Pluto will advance thermal model calculations to understand these zones. We also find substantial differences in albedo due to purely geophysical causes, including bifurcated albedos on Sputnik Planitia that may correspond to different episodes of activity, and the bifurcation of low-albedo material into the very dark terrain of Cthulhu, which may be a native substrate, and low-albedo material with normal reflectances of 0.20-0.25 that are associated with what appear to be dusty deposits from the Kuiper Belt or Pluto's haze layer. Alternatively this material could be akin to Cthulhu but with more higher albedo volatiles mixed in.    The surface of Charon is primarily water ice, while that of Pluto harbors the more volatile ices of nitrogen, CO and methane (Cruikshank et al. 2015; Grundy, 2016a). Pluto's larger mass was able to hold onto these transitory ices, to form a basis for seasonal transport of ice and an atmosphere, while Charon was only able to hold onto rock-like water ice.

One piece of data we lack is the measurement of Pluto and Charon at "true opposition", the geometry for which the geometric albedo is defined. The minimum in solar phase angles will be reached during Pluto's opposition in 2018. We have extrapolated the observations at small solar phase angles (~0.10°), which we observed in 2013, to obtain geometric albedos (Buratti et al., 2015). The opposition surge may be substantially larger if it is observed at even smaller solar phase angles (the minimum solar phase angle of 0.006° is reached in 2018). . Verbiscer et al. (2016) present measurements of the reflectance of Pluto and Charon at phase angles as small as 0.06° from HST observations obtained during the *New Horizons* epoch.

There are some obvious correlations between albedo and crater counts, e.g., Sputnik Planitia, the brightest region of Pluto, is free of craters. But the detailed picture is more complicated, with the lowest-albedo regions not the most crater-saturated (i.e., oldest) areas of Pluto (Stern et al., 2015; Robbins et al., 2016); instead, latitudinal patterns of local seasonal insolation and thermophysical structure may dominate (Binzel et al. 2016). Correlation of the albedo of crater ejecta deposits is another area of future study, particularly with respect to crater size. Increasingly larger craters excavate deeper into Pluto's crust and may uncover previous episodes of volatile deposition or reveal whether Pluto has a global low-albedo substrate.

One key result of this paper is that we have made a second connection between high (~unity) albedos from which the effects of viewing geometry have been eliminated, and geologic activity. The other example is Enceladus, with normal reflectances greater than 1.0 in some areas (Buratti and Veverka, 1984; Buratti, 1984; Buratti, 1988; Verbiscer et al.,1994). Moons embedded in the E-ring of Enceladus, such as Tethys, also have geometric albedos greater than unity (Verbiscer et al. 2007). Given that Eris has a geometric albedo of unity (Sicardy et al., 2011), it is also likely geologically active. The alternate theory offered by Sicardy et al. - that the high reflectivity is caused by the recent deposition of seasonally deposited frost - is less likely, as similar frost deposits on Triton (McEwen, 1990) and Pluto are not nearly as reflective. Furthermore, there is likely substantial amounts of dust in the Kuiper Belt (Stark, 2011) as well as native hydrocarbons (Simonelli et al. 1989) that would tend to darken frost. Clark (1981) showed that even a tiny amount of opaque material drastically lowers the albedo of an icy surface. Aerosols created in a hazy atmosphere, which is probably occurring on Pluto (Stern et al., 2015), may also serve as a darkening agent.

## Acknowledgements

Part of this research was carried out at the Jet Propulsion Laboratory, California Institute of Technology under contract to the National Aeronautics and Space Administration.

## References

Agnor, C. B., Hamilton, D. P., 2006. Neptune's capture of its moon Triton in a binary-planet gravitational encounter. *Nature* **441**, 192-194.

Bauer, J. et al., 2010. Direct Detection of Seasonal Changes on Triton with *Hubble Space Telescope*. *Ap. J. Lett*. **723**, L49-52.

Binzel, R. P. et al. 2016. Climate zones on Pluto and Charon. *Icarus*, accepted for publication.

Blackburn, D. G., Buratti, B. J., Rivera-Valentin, E. G., 2012. Exploring the impact of thermal segregation on Dione though a bolometric Bond albedo map. 43$^{rd}$ LPSC Conference, LPI Contribution # 1659.

Blackburn, D. G., Buratti, B. J., Ulrich, R., 2011. A bolometric Bond albedo map of Iapetus: Observations from Cassini VIMS and ISS and Voyager ISS. *Icarus* **212**, 329-338.

Bottke W. F., Vokrouhlický, D, Nesvorný, D., Moore J. M., 2013. Black rain: The burial of the Galilean satellites in irregular satellite debris. *Icarus* **223**, 775-795.

Buie, M. W., Grundy, W. M., Young, E. F., Young, L. A., Stern, S. A. 2010a. Pluto and Charon with the Hubble Space Telescope. I. Monitoring Global Change and Improved Surface Properties from Light Curves. *Astron. J.* **139**, 1117-1127.

Buie, M. W., Grundy, W. M., Young, E. F., Young, L. A., Stern, S. A. 2010b. Pluto and Charon with the Hubble Space Telescope. II. Resolving Changes on Pluto's Surface and a Map for Charon. *Astron. J.* **139**, 1128-1143.

Buratti, B. J., 1984. Voyager disk resolved photometry of the Saturnian satellites, *Icarus* **59**, 426-435.

Buratti, B. J., 1985. Application of a radiative transfer model to bright icy satellites. *Icarus* **61**, 208-217.

Buratti, B. J., 1988. Enceladus: Implications of its unusual photometric properties. *Icarus* **75**, 113-126.

Buratti, B. J., 1991. Ganymede and Callisto: Surface textural dichotomies and photometric analysis. *Icarus* **92**, 312-323.

Buratti, B. J., Golombek, M.,1990. Geologic implications of spectrophotometric measurements of Europa. *Icarus* **75**, 437-449.

Buratti, B.J., Mosher, J. A., 1995. The dark side of Iapetus: New evidence for an exogenous origin, *Icarus* **115**, 219-227.

Buratti, B. J., Mosher, J. A., Johnson, T. V., 1990. Albedo and color maps of the Saturnian satellites, *Icarus* **87**, 339-357.

Buratti, B. J., Veverka, J., 1983. Voyager photometry of Europa, *Icarus* **55**, 93-110.

Buratti, B. J., Veverka, J., 1984. Voyager photometry of Rhea, Dione, Tethys, Enceladus, and Mimas, *Icarus* **58**, 254-264.

Buratti, B. J., Veverka, J., 1985. The photometry of rough planetary surfaces: The importance of multiple scattering, *Icarus* **64**, 320-328.

Buratti, B. J., et al. 2003. Photometry of Pluto in the last decade and before: Evidence for volatile transport? *Icarus* **162**, 171-182.

Buratti, B. J. et al., 2011. Photometry of Triton 1992-2004: Surface volatile transport and discovery of a remarkable opposition surge. *Icarus* **212**, 835-846.

Buratti et al., 2015. Photometry of Pluto 2008-2014: Evidence of ongoing seasonal volatile transport and activity. *Ap. J. Lett*. **804**, L6-12.

Chandrasekhar, S. 1960. <u>Radiative Transfer</u>. Dover Press, New York.

Cheng, A. F. et al., 2008. Long-Range Reconnaissance Imager on *New Horizons*. *Space Science Reviews* **140,** 189-215.

Clark, R. N., 1981. The spectral reflectance of water-mineral mixtures at low temperatures. *J. Geophys. Res.* **86**, 3074-3086.

Cruikshank, D. P. et al., 2015. The surface compositions of Pluto and Charon. *Icarus* **246**, 82-92.

Fink, U., Disanti, M. A., 1988. The separate spectra of Pluto and its satellite Charon. *Astron. J.* **95**, 229-236.

Goguen, J. D., 1981. A theoretical and experimental investigation of the photometric functions of particulate surfaces. Ph.D. Thesis, Cornell University, Ithaca, NY.

Grundy, W. et al., 2016a. Surface compositions across Pluto and Charon. *Science* **351,** 1283.

Grundy, W. et al. 2016b. Formation of Charon's red polar caps. *Nature*, accepted for publication.

Hapke, B., 1981. Bidirectional reflectance spectroscopy. 1. Theory *J. Geophys. Res*. **86,** 3039-3054.

Hapke, B., 1984. Bidirectional reflectance spectroscopy. 3. Correction for macroscopic roughness. *Icarus* **59**, 41-59.

Hapke, B., 1986. Bidirectional reflectance spectroscopy. 4. The extinction coefficient and the opposition effect. *Icarus* **67**, 264-280.

Hapke, B., 1990. Coherent backscatter and the radar characteristics of our planet satellites. *Icarus* **88**, 407-417.

Hapke, B., 2012. Theory of Reflectance and Emittance Spectroscopy, Second Edition. Cambridge University Press, Cambridge and New York.

Helfenstein, P., Veverka, J., Thomas, P. C., 1988. Uranus satellites: Hapke parameters from Voyager disk integrated photometry. *Icarus* **215,** 83-100.

Helfenstein, P., Shepard, M. K. 2011. Testing the Hapke photometric model: Improved inversion and the porosity correction. *Icarus* **74**, 231

Hicks, M. D., Buratti, B. J., 2004. Spectroscopy and filter photometry of Triton from 1997-1999: Evidence for dramatic global changes? *Icarus* **171,** 210-218.

Hillier, J., Helfenstein, P., Verbiscer, A., Veverka, J., Brown, R. H., Goguen, J., Johnson, T. V., 1990. Voyager disk-integrated photometry of Triton. *Science* **250,** 419-421.

Horak, H., 1950. Diffuse reflection by planetary atmospheres. *Ap. J.* **112**, 445-463.

Irvine, W. M., 1966. The shadowing effect in diffuse radiation. *J. Geophys. Res.* **71**, 2931-2937.

Johnson, T. V. et al., 1983. Global multispectral mosaics of the icy Galilean satellites. *J. Geophys. Res*. **88,** 5789-5805.

Lane, A. P., Irvine, W. M., 1973. Monochromatic phase curves and albedos for the lunar disk. *Astron. J.* **78**, 267-277.

Li. J.-Y. et al., 2013. Global photometric properties of Asteroid (4) Vesta observed with Dawn Framing Camera. *Icarus* **226**, 1252-1274.

Marcialis, R., 1988. A two-spot albedo model for the surface of Pluto. *Astron. J.* **95**, 941-947.

McEwen, A.S., 1990. Global color and albedo variations on Triton. *J. Geophys. Res*. **17**, 1765-1768.

Minnaert, M., 1961. "Photometry of the Moon." In The Solar System III. Planets and Satellites. (G.P. Kuiper and B. M. Middlehurst, eds.), University of Chicago Press, Chicago.

Robbins et al., 2016. Craters of the Pluto-Charon System, Submitted to *Icarus*.

Sawyer, S.R., Barker, E. S., Cochran, A. L., Cochran, W. D., 1987. Spectrophotometry of Pluto-Charon mutual events - Individual spectra of Pluto and Charon. *Science* **238**, 1560-1563.

Schenk, P. et al., 2011. Plasma, plumes and rings: Saturn system dynamics as recorded in global color patterns on its midsize icy satellites. *Icarus* **211**, 740-757.

Shepard, M. K., Helfenstein, P., 2007. A test of the Hapke photometric model. Theory *J. Geophys. Res*. **112,** E03001.

Shepard, M. K., Helfenstein, P., 2011. A laboratory study of the bidirectional reflectance from particulate Theory. *Icarus* **215**, 526-533.

Shkuratov, Y. G., Grynko, Y. S., 2005. Light scattering by media composed of semitransparent particles of different shapes in ray optics approximation: consequences for spectroscopy, photometry, and polarimetry of planetary regoliths. *Icarus* **173**, 16-28.

Sicardy, B. et al., 2011. A Pluto-like radius and a high albedo for the dwarf planet Eris from an occultation. *Nature* **478**, 493-496.

Simonelli, D. P., Pollack, J. B., McKay, C. P., Reynolds, R. T., Summers, A. L., 1989. The carbon budget in the outer solar nebula. *Icarus* **82**, 1-35.

Simonelli, D. P., Kay, J., Adinolfi, D., Veverka, J., Thomas, P. C., Helfenstein, P., 1999. Phoebe: Albedo Map and Photometric Properties. *Icarus* **138**, 249-258.

Squyres S., Veverka, J., 1981. Voyager photometry of surface features on Ganymede and Callisto. *Icarus* **46**, 137-155.

Squyres S., Veverka, J., 1982. Variation of Albedo with Solar Incidence Angle on Planetary Surfaces. *Icarus* **50**, 115-122.

Stark, C. 2011. The Transit Light Curve of an Exozodiacal Dust Cloud. *Astron. J*. **142**, article id. 123, 11 pp.

Stern, S. A., Trafton, L. M., Gladstone, G. R., 1998. Why is Pluto bright? Implications of the albedo and lightcurve behavior of Pluto. Icarus 75, 485-498

Stern, S. A., Buie, W. M., Trafton, L., 1997. HST high-resolution maps and images of Pluto. *Astron. J*. **113**, 827-843.

Stern, S. A. et al., 2015. The Pluto System: Initial Results from its Exploration by New Horizons. *Science* **350**, 292-300.

Spencer, J. R., Denk, T., 2010. Formation of Iapetus' Extreme Albedo Dichotomy by Exogenically Triggered Thermal Ice Migration. *Science* **327**, 432-435.Verbiscer, A. J., French, R. G., McGhee, C. A., 2005. The opposition surge of Enceladus: HST observations 338-1022 nm. *Icarus* **173**, 66-83.

Verbiscer, A. J., Veverka, J. 1994. A photometric study of Enceladus. *Icarus* **110**, 155-164.

Verbiscer, A. J., French, R. G., Showalter, M., Helfenstein, P., 2007. Enceladus: Cosmic graffiti artist caught in the act. *Science* 315, 815.

Verbiscer, A. J., Skrutskie, M. F., Hamilton, D. P., 2009. Saturn's largest ring. *Nature* **461**, 1098-1100.

Verbiscer, A. J. et al. 2016. The Pluto System at Small Phase Angles. Talk to be presented at DPS/EPSC meeting Oct. 17-21, 2016, Pasadena, CA.

Young, E. F., Galdamez, K., Buie, M. W., Binzel, R. P., Tholen, D., 1999. Mapping the Variegated Surface of Pluto. *The Astronomical Journal* **117**, 1063-1076.

Table 1 - Full Disk Images used for albedo maps (all angles are in degrees)

| LORRI Image # | UTC July 2015 | Exp (s) | Target | α | Lat Subspacecraft | Long Subspacecraft | Lat Subsolar | Long Subsolar | Res (km/pixel) | Range (km) |
|---|---|---|---|---|---|---|---|---|---|---|
| LOR_0298615084 | 07-22:46:05.7 | 0.10 | P | 15.08 | 43.11 | 155.31 | 51.53 | 136.73 | 38.67 | 7792553.7 |
| LOR_0298615084 | 07-22:46:05.7 | 0.10 | C | 15.08 | 43.03 | 335.39 | 51.53 | 316.73 | 38.74 | 7805416.1 |
| LOR_0298721714 | 09-04:23:15.8 | 0.10 | P | 15.11 | 43.09 | 85.75 | 51.54 | 67.15 | 31.39 | 6324328.1 |
| LOR_0298721714 | 09-04:23:15.8 | 0.10 | C | 15.11 | 43.10 | 266.00 | 51.54 | 247.16 | 31.38 | 6322923.3 |
| LOR_0298787094 | 09-22:32:55.8 | 0.10 | P | 15.15 | 43.06 | 43.13 | 51.54 | 24.50 | 26.92 | 5424146.7 |
| LOR_0298787094 | 09- 22:32:55.8 | 0.10 | C | 15.15 | 43.17 | 223.32 | 51.54 | 204.50 | 26.87 | 5413459.7 |
| LOR_0298787344 | 09-22:37:05.8 | 0.10 | P | 15.15 | 43.06 | 42.97 | 51.54 | 24.33 | 26.90 | 5420703.7 |
| LOR_0298787344 | 09-22:37:05.8 | 0.10 | C | 15.15 | 43.17 | 223.16 | 51.54 | 204.33 | 26.85 | 5409989.6 |
| LOR_0298893504 | 11- 04:06:25.8 | 0.10 | P | 15.24 | 43.01 | 333.82 | 51.54 | 315.07 | 19.64 | 3957621.4 |
| LOR_0298893754 | 11-04:10:35.8 | 0.10 | C | 15.24 | 43.19 | 153.47 | 51.54 | 134.91 | 19.56 | 3941522.6 |

| LORRI Image | Image midtime | Exp (s) | Target | α | | | Subsolar Lat | | | Range (km) |
|---|---|---|---|---|---|---|---|---|---|---|
| LOR_0298959350 | 11-22:23:51.8 | 0.10 | P | 15.31 | 42.98 | 290.96 | 51.55 | 272.11 | 15.13 | 3049149.9 |
| LOR_0298959599 | 11-22:28:00.8 | 0.10 | C | 15.33 | 43.06 | 110.32 | 51.55 | 91.94 | 15.09 | 3041035.2 |
| LOR_0298959629 | 11-22:28:30.8 | 0.10 | C | 15.33 | 43.06 | 110.30 | 51.55 | 91.93 | 15.09 | 3040625.8 |
| LOR_0298996724 | 12-08:46:45.8 | 0.10 | P | 15.39 | 42.95 | 266.65 | 51.55 | 247.72 | 12.57 | 2533343.9 |
| LOR_0298996974 | 12-08:50:55.8 | 0.10 | C | 15.39 | 42.93 | 85.89 | 51.55 | 67.56 | 12.56 | 2531227.9 |
| LOR_0298997004 | 12-08:51:25.8 | 0.10 | C | 15.39 | 42.93 | 85.87 | 51.55 | 67.54 | 12.56 | 2530818.8 |
| LOR_0299075349 | 13-06:37:10.7 | 0.15 | C | 15.64 | 42.40 | 35.04 | 51.55 | 16.42 | 7.25 | 1460446.9 |
| LOR_0299123689 | 13-20:02:50.8 | 0.10 | P | 16.07 | 42.55 | 184.57 | 51.55 | 164.88 | 3.88 | 781930.7 |
| LOR_0299124574 | 13-20:17:35.8 | 0.10 | P | 16.08 | 42.54 | 84.01 | 51.55 | 164.31 | 3.81 | 769733.3 |
| LOR_0299147641 | 14-02:42:02.7 | 0.15 | C | 16.85 | 40.50 | 350.33 | 51.55 | 329.26 | 2.31 | 466350.5 |

Target: P=Pluto
C=Charon

α = solar phase angle

Table 2- Close-encounter Images used for high-resolution studies

| LORRI Image | Image midtime (UTC) 2015-07-14 | Exp (s) | α | LORRI boresight incident, emission angles | Subspacecraft Lat | Subspacecraft Long | Sub-solar Long | Res (km/pixel) | Range (km) |
|---|---|---|---|---|---|---|---|---|---|
| LOR_0299168135 | 08:23:37 | 0.15 | 19.64 | 32,21 | 40.15 | 159.15 | 135.88 | 0.84 | 168784 |
| LOR_0299177051 | 10:52:13 | 0.05 | 31.59 | 32,22 | 31.65 | 163.40 | 130.07 | 0.23 | 47379 |
| LOR_0299177087 | 10:52:49 | 0.05 | 31.55 | 38,19 | 31.52 | 163.51 | 130.04 | 0.23 | 46878 |
| LOR_0299177195 | 10:54:37 | 0.05 | 31.47 | 63,35 | 31.13 | 163.84 | 129.97 | 0.23 | 45605 |
| LOR_0299168184 | 08:24:26 | 0.15 | 19.42 | 68,50 | 40.14 | 159.13 | 135.85 | 0.83 | 168460 |
| LOR_0299168727 | 08:33:29 | 0.15 | 24.27 | 56,74 | 36.16 | 342.00 | 315.50 | 0.87 | 175689 |

All images are of Pluto except the last one, which is of Charon
Subsolar latitude was 51.55
α = Solar phase angle

Table 3: Albedo Variations on Icy Bodies

| Object | Maxiumum | Minimum | Source |
|---|---|---|---|
| Pluto | 1.0 | 0.08 | This study |
| Charon | 0.73 | 0.11 | This study |
| Iapetus | 0.70 | 0.02 | Buratti et al. 1990 |
| Europa | 0.85 | 0.55 | Buratti and Golombek, 1988 |
| Triton | 0.90 | 0.62 | McEwen, 1990 |
| Enceladus | 1.4 | 0.90 | Verbiscer and Veverka, 1994; Verbiscer et al., 2007 |
| Callisto | 0.70 | 0.14 | Squyres et al., 1981 |
| Dione | 0.60 | 0.25 | Buratti, 1984 |
| Phoebe | 0.07 | 0.13 | Simonelli et al., 1999 |

**Figure Captions**

Figure 1. The solar phase curves of Pluto and Charon compared with other Solar System objects. The existing curves are based on Buratti and Veverka, 1984; Buratti, 1991; Hillier et al., 1990, and Lane and Irvine, 1973.

Figure 2. A map of normal reflectance of Pluto. Images used in the map are listed in Table 1.

Figure 3. Three mosaics of Charon with images processed using a range of values of A for equation 1. Clearly a lunar like value of 1.0 does not adequately remove geometric effects from the images. Both limb-brightening and limb-darkening remain in the images. The best fits are provided by A=0.7, similar to that for Pluto.

Figure 4. A map of the normal reflectance of Charon. Images used in the map are listed in Table 1.

Figure 5. A preliminary map of the Bond albedo of Pluto from LORRI images.

Figure 6. A preliminary map of the Bond albedo of Charon from LORRI images.

Figure 7. A map of Pluto showing the locations of the images listed in Table 2. The numbers on the maps are the last three digits of the LORRI picture numbers.

Figure 8. Albedo maps and histograms of the LORRI images plotted in Figure 7.

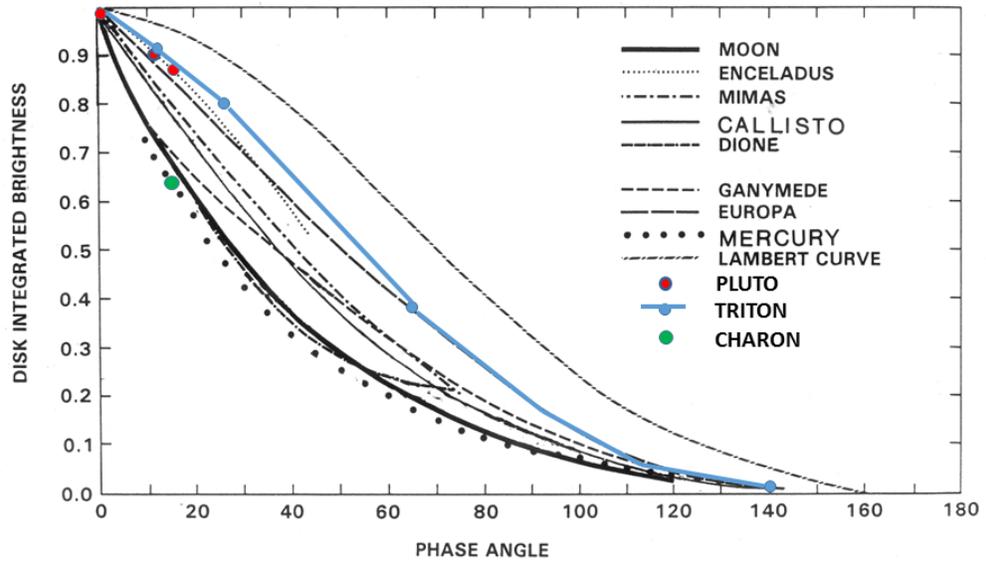

Figure 1

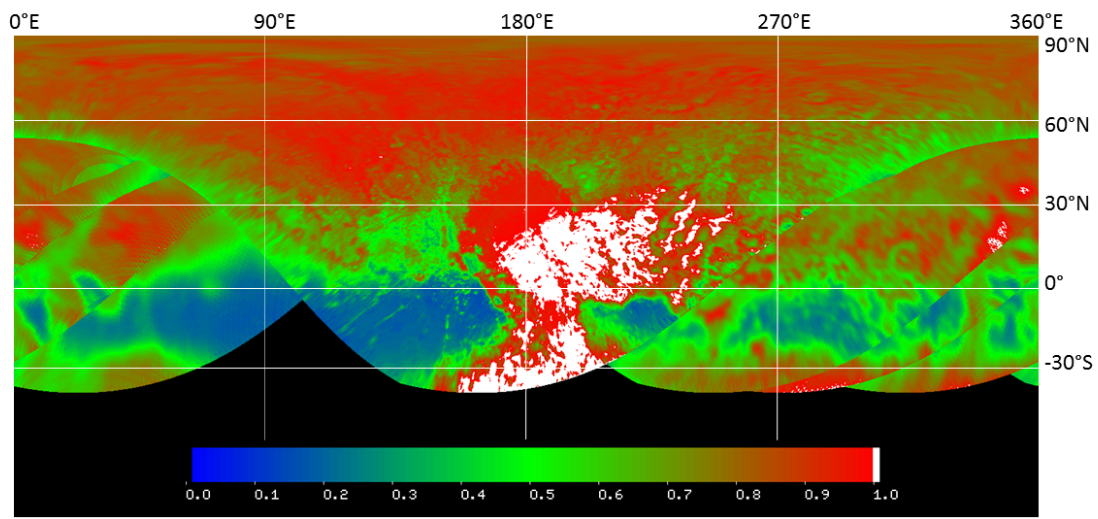

Figure 2

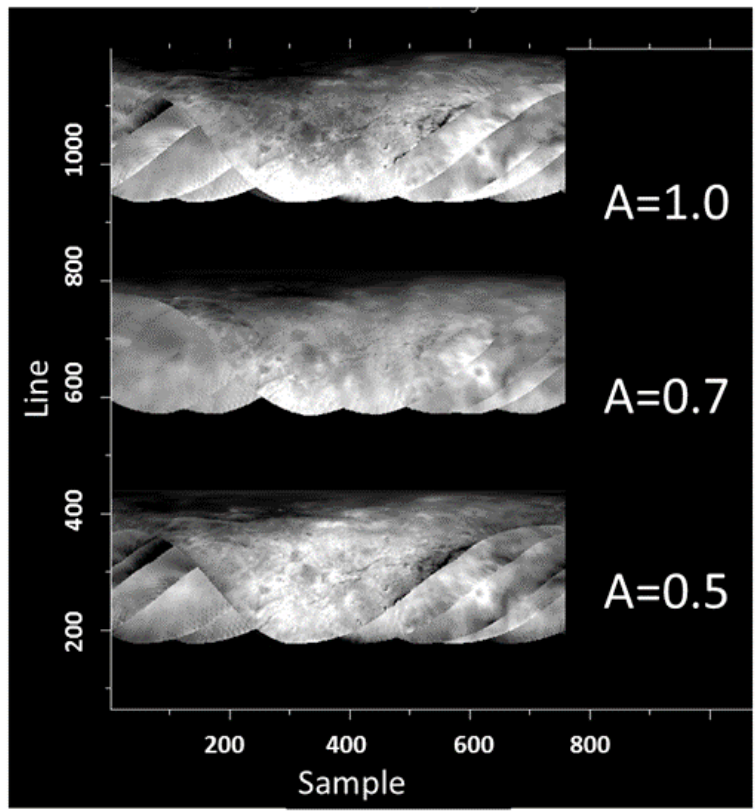

Figure 3

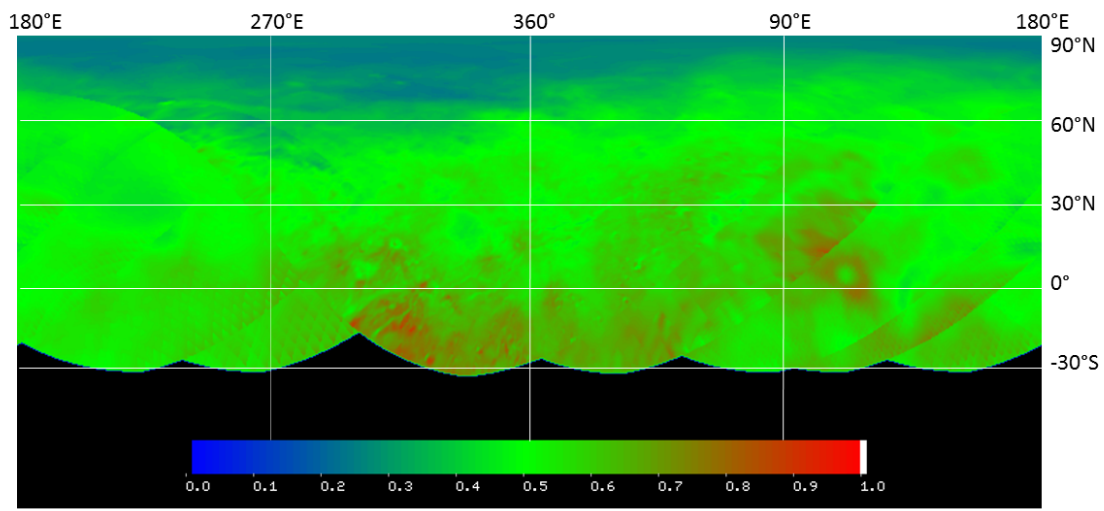

Figure 4

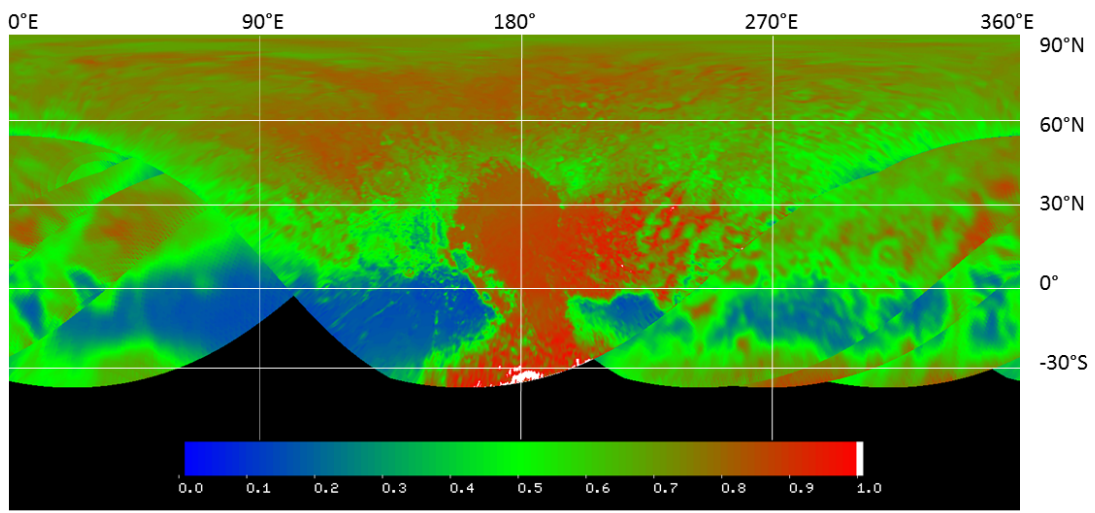

Figure 5

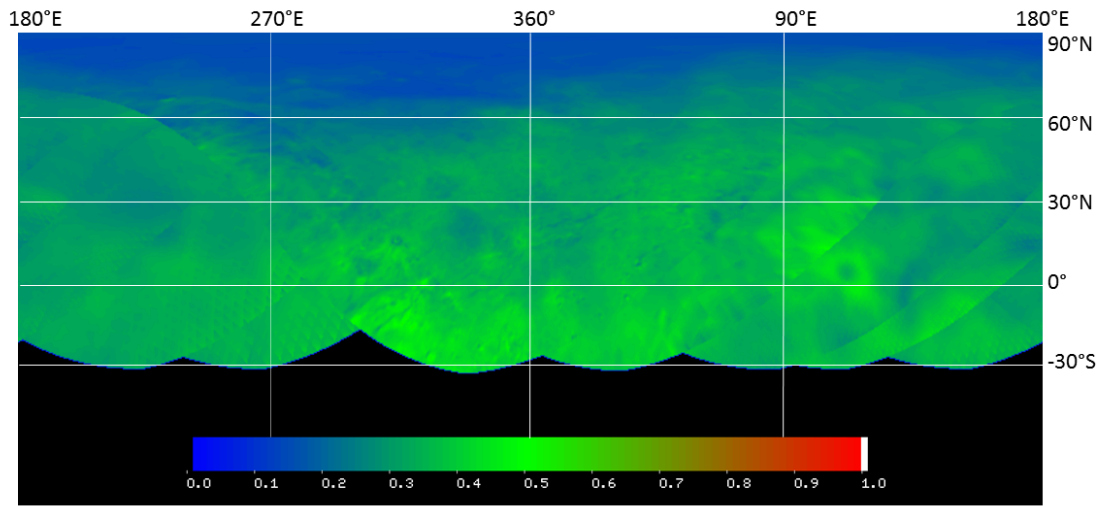

Figure 6

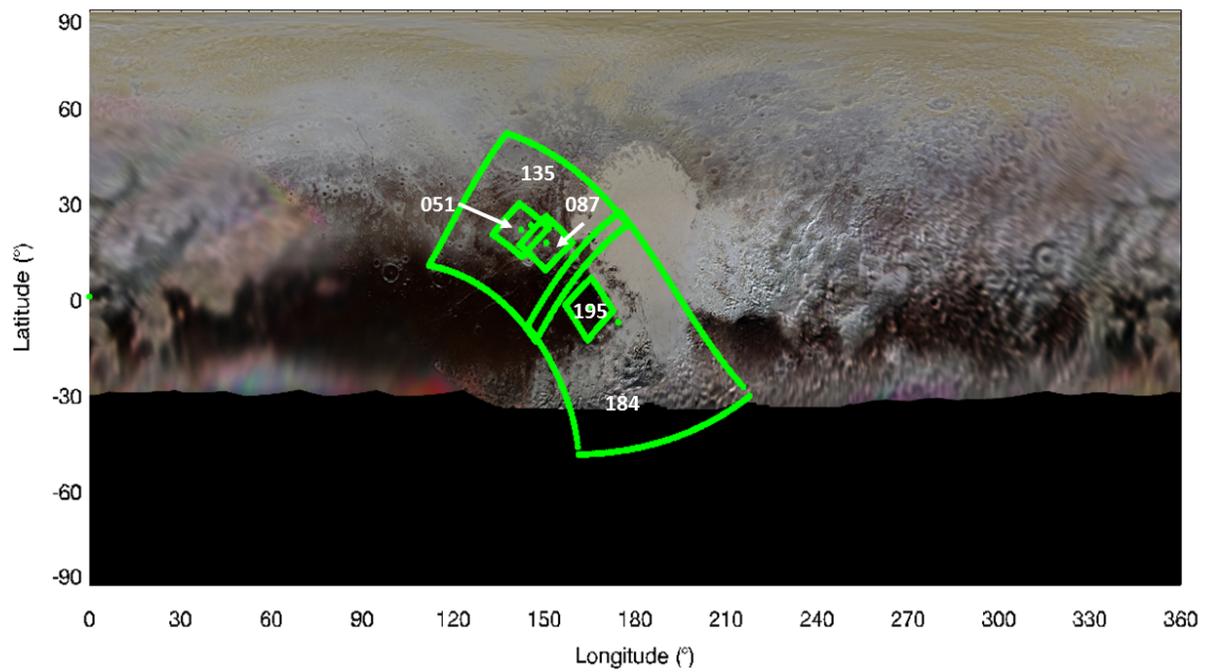

Figure 7

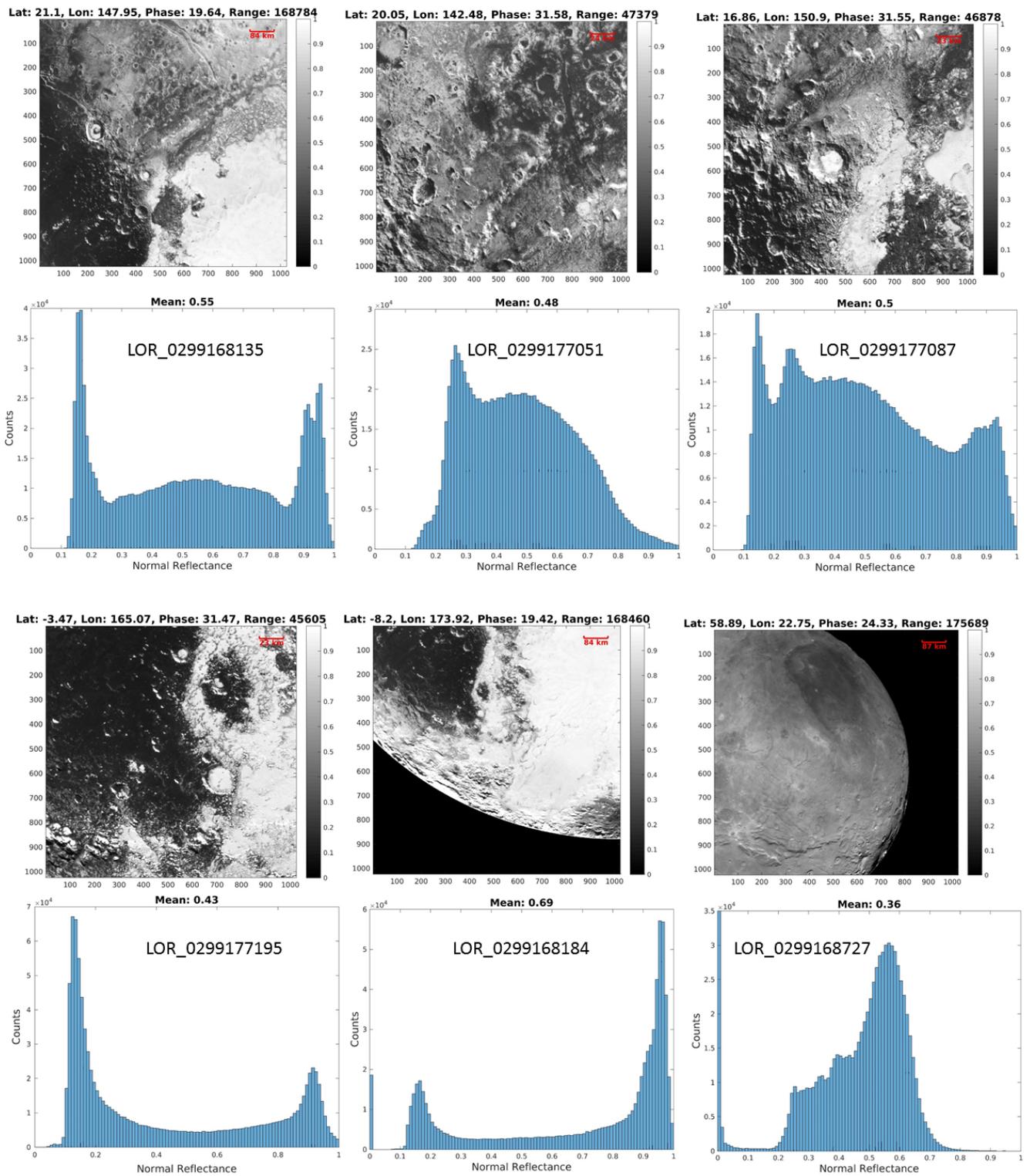

Figure 8